\documentclass[11pt]{article}

\usepackage{geometry}                
\usepackage{graphicx,epsfig,framed,cite}

\usepackage[pdftex,colorlinks=false]{hyperref}

\begin{document}

\def\describeRec{Provide a clear,
explicit description of the analysis in publications.  In particular,
the most crucial information such as basic object definitions and
event selection should be clearly displayed in the publications,
preferably in tabular form, and kinematic variables utilised should be
unambiguously defined.  Further information necessary to reproduce the
analysis should be provided, as soon as it becomes available for
release, on a suitable common platform.}
\def\databaseRec{The community should identify, develop and adopt a 
common platform to store analysis databases, collecting object
definitions, cuts, and all other information, including
well-encapsulated functions, necessary to reproduce or use the
results of the analyses, and as required by other recommendations.}
\def\efficiencyRec{Provide histograms or functional forms of efficiency maps 
wherever possible in the auxiliary information, along with precise definitions 
of the efficiencies, and preferably provide them in standard electronic forms 
that can easily be interfaced with simulation or analysis software.}
\def\simulatorRec{The community should take responsibility for providing, 
validating and maintaing a simplified simulation code for public use, reproducing 
the basic response of the LHC detectors. The validation and tuning of this tool 
should be based on comparisons with actual performance plots, and/or other 
inputs, made available by the experiments along the lines of Recommendation~2a. 
Limits of validity should be investigated and clearly documented.}
\def\analysisNumbersRec{Provide all crucial numbers regarding the results of the analysis, 
preferably in tabulated form in the publication itself. Further relevant information, like fit 
functions or distributions, should be provided as auxiliary material.}
\def\multiBinRec{For multi-bin results, provide an ensemble of sets of the numbers $B$,
$\delta B$, $\cal{L}$, $\delta\cal{L}$, $Q$, $k$, etc in the auxiliary information.
These would be created by sampling from the various experiment-specific systematic effects,
such as the jet energy scale, jet energy resolution, etc. 
Results should be quoted without inclusion of systematic/theoretical uncertainties
external to the experiment.} 
\def\likelihoodRec{When feasible, 
provide a mathematical description of the \underline{final} likelihood 
function in which experimental data and parameters are clearly distinguished, 
either in the publication or the auxiliary information. 
Limits of validity should always be clearly specified.}
\def\roostatsRec{Additionally provide a digitized implementation of the
likelihood that is consistent with the mathematical description.}
\def\interpretRec{In the interpretation of experimental results,  
preferably provide the final likelihood function (following Recommendations~3b/3c). 
When this is not possible or desirable, provide a grid of confidence levels over the parameter space.  
The expected constraints should be given in addition to the observed ones, and 
whatever sensitivity measure is applied must be precisely defined. 
Modeling of the acceptance needs to be precisely described.}
\def\higgsRec{For Higgs searches, provide all relevant information on a channel-by-channel 
basis for both production and decay processes.}
\def\designRec{When relevant, 
design analyses and signal regions that are based on  disjoint sets of events.}

\newcounter{recom}
\setcounter{recom}{0}

\vspace*{1cm}

\noindent
{\Large\bf Searches for New Physics: Les Houches Recommendations\\[1mm] 
for the Presentation of LHC Results}

\vspace{8mm}

\noindent
S.~Kraml$^1$, 
B.C.~Allanach$^2$, 
M.~Mangano$^3$, 
H.B.~Prosper$^4$, 
S.~Sekmen$^{3,4}$ (editors),\\
C.~Balazs$^5$, 
A.~Barr$^6$,
P.~Bechtle$^7$, 
G.~Belanger$^8$, 
A.~Belyaev$^{9,10}$,
K.~Benslama$^{11}$, 
M.~Campanelli$^{12}$, 
K.~Cranmer$^{13}$, 
A.~De~Roeck$^3$, 
M.J.~Dolan$^{14}$, 
T.~Eifert$^{15}$, 
J.R.~Ellis$^{16,3}$, 
M.~Felcini$^{17}$, 
B.~Fuks$^{18}$, 
D.~Guadagnoli$^{8,19}$,
J.F.~Gunion$^{20}$, 
S.~Heinemeyer$^{17}$, 
J.~Hewett$^{15}$, 
A.~Ismail$^{15}$, 
M.~Kadastik$^{21}$,
M.~Kr\"amer$^{22}$,
J.~Lykken$^{23}$
F.~Mahmoudi$^{3,24}$,
S.P.~Martin$^{25,26,27}$,
T.~Rizzo$^{15}$, 
T.~Robens$^{28}$, 
M.~Tytgat$^{29}$, 
A.~Weiler$^{30}$\\

\vspace{8mm}

\hrule
\begin{abstract}
We present a set of recommendations for the presentation of LHC   
results on searches for new physics, which are aimed at providing a more efficient flow of scientific information between the experimental collaborations and the rest of the high energy physics community, and at facilitating the interpretation of the results in a wide class of models. Implementing these recommendations would aid the full exploitation of the physics potential of the LHC.  
\end{abstract}
\hrule

\vspace*{12mm}

\noindent{\it 
$^1$\,Laboratoire de Physique Subatomique et de Cosmologie, UJF Grenoble 1, 
CNRS/IN2P3, INPG, 53~Avenue des Martyrs, F-38026 Grenoble, France\\
$^2$\,DAMTP, CMS, University of Cambridge, Wilberforce Road, Cambridge, CB3 0WA,
United Kingdom\\
$^3$\,Physics Department, CERN, CH 1211 Geneva 23, Switzerland\\
$^4$\,Department of Physics, Florida State University, Tallahassee, Florida 32306, USA\\
$^5$\,Monash University, School of Physics, Melbourne Victoria 3800 Australia\\
$^6$ Department of Physics, Keble Road, Oxford, OX1 3RH, United Kingdom\\
$^7$\,Universit"at Bonn, Physikalisches Institut, Nussallee 12, 53115 Bonn, Germany\\
$^8$\,LAPTH, Universit\'e de Savoie, CNRS, B.P.110, F-74941 Annecy-le-Vieux Cedex, France\\
$^9$\,NExT Institute: School of Physics and Astronomy, Univ. of Southampton, UK\\
$^{10}$\,Particle Physics Department, Rutherford Appleton Laboratory, UK\\
$^{11}$\,Physics Department, University of Regina,3737 Wascana Parkway, Regina--SK, Canada\\
$^{12}$\,University College London, Gower Street, WC1B 6BT London, UK\\
$^{13}$\,Center for Cosmology and Particle Physics, New York University, New York, NY 10003, USA\\
$^{14}$\,Institute for Particle Physics Phenomenology, University of Durham, Durham DH1 3LE, UK\\
$^{15}$\,SLAC National Accelerator Laboratory, 2575 Sand Hill Rd, Menlo Park, CA 94025, USA\\
$^{16}$\,Theoretical Particle Physics and Cosmology Group, Department of Physics, King's College London, London WC2R 2LS, UK\\
$^{17}$\,Instituto de F\'{\i}sica de Cantabria (IFCA), CSIC-Universidad de Cantabria, E--39005 Santander, Spain\\ 
$^{18}$\,Institut Pluridisciplinaire Hubert Curien/D\'epartement Recherches Subatomiques, Universit\'e de Strasbourg/CNRS-IN2P3, 23 Rue du Loess, F-67037 Strasbourg, France\\
$^{19}$\,Laboratoire de Physique Th\'eorique, Universit\'e Paris-Sud, Centre d'Orsay,
F-91405 Orsay-Cedex, France\\
$^{20}$\,Department of Physics, University of California at Davis, Davis CA, USA\\
$^{21}$\,National Institute of Chemical Physics and Biophysics (NICPB), Tallinn, Estonia\\
$^{22}$\,Institute for Theoretical Particle Physics and Cosmology, RWTH Aachen University, D-52056 Aachen, Germany\\
$^{23}$\,Fermi National Accelerator Laboratory, P.O. Box 500, Batavia, IL 60510\\
$^{24}$\,Clermont Universit\'e, Universit\'e Blaise Pascal, CNRS/IN2P3, LPC, BP 10448, 63000 Clermont-Ferrand, France\\
$^{25}$\,Department of Physics, Northern Illinois University, DeKalb IL 60115, USA\\
$^{26}$\,Fermi National Accelerator Laboratory, P.O. Box 500, Batavia IL 60510, USA\\
$^{27}$\,Kavli Institute for Theoretical Physics, University of California, Santa Barbara CA 93106, USA\\
$^{28}$\,IKTP, TU Dresden, Zellescher Weg 19, 01069 Dresden, Germany\\
$^{29}$\,Ghent University, Dept. of Physics and Astronomy, Proeftuinstraat 86,
B-9000 Gent, Belgium\\
$^{30}$\,Deutsches Elektronen-Synchrotron DESY, Notkestra{\ss}e 85, D-22607 Hamburg, Germany\\
}

\clearpage

\section{Introduction}

The LHC has very successfully begun to explore the TeV energy scale, 
and will be the energy frontier machine for the foreseeable future. Everyone who has had
a hand in bringing this scientific and technological
marvel to fruition deserves considerable credit and our thanks: 
the physicists and engineers who
conceived, designed, and built it; those who operate the machine and its experiments;
those who produce experimental results; those who try to understand them, and the public 
and its representatives whose generous support has enabled all this to happen. 

The LHC was designed as a machine of discovery.
There are high hopes that groundbreaking discoveries will indeed occur and 
shed light on  
electroweak symmetry breaking (be it via the Higgs mechanism or some other new
dynamics) and new physics beyond the Standard Model (SM) of electroweak and strong
interactions.  It is of highest priority to our community to
exploit fully the physics potential of the LHC.
One aspect of this exploitation is the interpretation of LHC results in the contexts 
of different models of new physics. This is crucial if we are to unravel the correct 
new physics model, determine its parameters, and move beyond the SM. 

The ATLAS and CMS collaborations are providing detailed
experimental results~\cite{Atlastwiki,Cmstwiki} of searches in many different channels. 
They are also providing interpretations in terms of popular models, such as the 
CMSSM\footnote{Constrained Minimal Supersymmetric Standard Model, see e.g.~\cite{AbdusSalam:2011fc}.}, 
or in terms of Simplified Models\footnote{Simplified Models are designed as an effective-Lagrangian description 
of a small number of accessible new particles. This approach has a long heritage; for a recent paper
advocating it see e.g.~\cite{Alves:2011wf}.}. 
These results are being used  
to test as large a variety of beyond-the-SM (BSM) scenarios as possible. 
For example, the searches for supersymmetry (SUSY), 
including~\cite{Chatrchyan:2011zy,Aad:2011cwa,Aad:2011ib,ATLAS:2011ad,CMS:SS,CMS:OS},
were interpreted in a number of 
different SUSY-breaking schemes, see e.g.\ \cite{Farina:2011bh,Buchmueller:2011sw,Fowlie:2011mb,Allanach:2011qr,Kats:2011qh,Grellscheid:2011ij}, 
as well as in the weak-scale ``phenomenological'' MSSM~\cite{Sekmen:2011cz,Arbey:2011un}. 
The sensitivity to light stops was investigated 
in~\cite{Papucci:2011wy,Bi:2011ha,Desai:2011th}, while implications of
compressed SUSY spectra were analyzed in~\cite{LeCompte:2011fh}. 
Interpretations were also made for non-SUSY models, for instance for the minimal 
universal extra dimension (UED) model in~\cite{Chang:2011aa}.  
Similar non-collaboration efforts to interpret Higgs search results~\cite{atlas:2012si,Chatrchyan:2012tx} 
in a large variety of BSM scenarios are also underway.  
These examples illustrate the community's interest in the LHC
experimental results---interest that will surely grow as results become more 
comprehensive and readily available.

A systematic way of presenting LHC results will
also greatly facilitate the comparison and combination of analyses
within and across the LHC collaborations, as well as the assessment of
the physics potential of future facilities. Furthermore, agreement on a set
of recommendations and their implementation would be
a further step towards a more comprehensive approach to the storage,
persistence and future use of LHC results.

In this report, we therefore propose 
a set of recommendations for the presentation of LHC results
aimed at maximizing its scientific return.  Many of the
experimental publications already implement several of the basic
recommendations we make. But, as we shall see,  our
recommendations go substantially beyond current practice. Our wish is to stimulate discussions among the whole community and work towards an agreement on
a common standard for the presentation of results. The goal is 
to help the community make the most of an extraordinary scientific
opportunity.

\section{Guiding Principles}

As mentioned, it is our purpose here to formulate a set of
recommendations that could act as guidelines for the documentation and
use of the LHC results, in a form that would be most useful to the
community at large, and that would help to maximize the scientific
output of the LHC.  Our recommendations are intended to respect the intellectual 
property rights of the collaborations and be concrete, practical and
clear, as well as not being burdensome for the scientists performing
the experimental analyses.  By and large, several of these
recommendations are rather obvious, and are already implemented in the
publications of the LHC experiments. On the other hand, others are
more ambitious. We present them here to stimulate further discussion
between the experiments and the community, in the hope that they could
be eventually adopted as part of the common practice.  

With this in
mind, our recommendations are guided by the following principles:
\begin{itemize}
\item \emph{What} has been observed should be clear to a non-collaboration
  colleague. 
\item \emph{How} it has been observed should be clear to a non-collaboration
  colleague. 
\item An interested non-collaboration colleague should be able to use and (re-)interpret results without the need to take up the time of collaboration insiders.
\end{itemize}
The latter principle implies that all ingredients (e.g., data, experimental systematics, cuts, procedures and so forth)
in the analysis should be completely 
and unambiguously specified. We are not, of course, arguing that
scientific discourse either within a collaboration or between
collaboration members and those outside should be curtailed. On the contrary, this is vital to maintain the intellectual vibrancy of
the field, and our suggestions are intended to make this more efficient and to reduce the burden on collaboration members.

To this aim, we think it useful to distinguish between experimental results and their interpretation. 
We suggest that the term \emph{experimental result} be used exclusively to
mean the empirical outcome, such as an event count or the measurement of some
physical quantity. The experimental results themselves should be 
independent of any hypothesized new physics model. The term \emph{interpretation} is the
act of  comparing the experimental results to model predictions.  
If full details of the experimental results employed for a given interpretation 
are readily available, their interpretation in different new
physics models becomes possible.  While the design of an analysis may have
been inspired or guided by a specific physics model, ideally, the experimental results themselves
should be independent of any such model so that the results can, in principle, be subject to different
interpretations.

We also emphasize that it is important for the legacy of the LHC that its 
experimental results can be used in the future, even after the LHC has shut
down and the collaborations have been disbanded. 
A coherent strategy for data preservation and re-use is discussed in~\cite{Kogler:2011sn}.

These considerations, along with the principles listed above, have guided 
the recommendations in this document.  We note again that many of these recommendations are already implemented in experimental analyses and publications: we hope that this document will facilitate discussions of---and serve as a guide for---best practice. 

\section{Recommendations}

In the following we discuss our recommendations, which we present in four broad categories:
analysis description, detector modeling, analysis dissemination and analysis design. 
Moreover, we include some recommendations regarding the interpretation of the results. 
Where appropriate, we split our recommendations into options: 
\begin{description}
\item[] {\bf (a)} ``crucial" recommendations, defined as actions that we believe should be undertaken immediately, and 
\item[] {\bf (b), (c)} ``desirable steps'', i.e.\ actions that would help, but whose implementation is recognized as requiring major efforts and a longer timescale. 
\end{description}
Recommendations without such sub-division are understood as  ``crucial".

\subsection{Analysis Description \label{sec:analysis}}

As noted above, our guiding principle is that an interested
non-collaboration colleague should be provided with all of the
necessary information that is needed to use published results without
having to consult collaboration insiders (although it may be wise to
do so anyway).  We thus recommend that the experimental publications
contain a description of the analyses as clear and explicit as
possible.  
Basic object definitions --- for example, what constitutes an
isolated electron --- should be specified, so that the analysis may be
reliably reproduced.  Definitions of the important variables for the
analysis should be precisely stated, because different definitions or
conventions may exist for selection variables such as 
$E_T^{\rm miss}$, $m_{\rm eff}$, $H_T$, $M_{T2}$, $\alpha_T$.  

It is crucial that the analysis description provides sufficient information to
validate an implementation of the analysis by users.  In this regard,
providing {\em cutflows}, i.e., the number of events obtained after each stage of the event
selection for a given data or Monte Carlo (MC) event set, would provide valuable assistance.
Since non-collaboration colleagues do not have access to the experimental data, nor
the MC event set simulated with an official collaboration detector simulation, they
do not have direct means to perform an exact, one-to-one synchronization and
validation.  It would help substantially to adopt the common practice of
providing cutflows for a set of MC events, as well as experimental data, for a physics
process that can be easily reproduced. 
Relevant information defining these MC events,
e.g.\ the underlying physics model and processes, and the details of tools used
in pre-detector event generation, including version information, should be specified.  
We note that guidelines
for using event generators already exist  in {\tt MCNET}~\cite{MCNET}, see also \cite{Butterworth:2010ym},  
and we re-emphasize to adopt them. 

Access to all this necessary information will be facilitated if it is tabulated, rather
then described in the text. If limits on publication length do not allow the inclusion 
of all relevant information in the publication itself, the remaining details could be 
provided as {\em auxiliary information} alongside the publication. 
It would further greatly help to provide the relevant 
information in figures (coordinates of points in a graph, events in a
histogram, etc.) in a digital form that is easily readable, e.g., as lists of
numbers, as self-contained functions or as {\tt ROOT} objects, etc.\  
We thus summarize \clearpage

\addtocounter{recom}{1}
\begin{quote}
  {\bf Recommendation~\arabic{recom}a:} \emph{\describeRec} 
\end{quote}

We note that it is already common practice in the LHC experiments to provide useful 
auxiliary information for the longer papers\footnote{It is understood that, 
in order not to delay the publication while all the supplementary information is being prepared, 
not all of this information may be available immediately with the release of a paper, 
in particular when shorter articles such as Letters are concerned.}, e.g.,
in {\tt Rivet}~\cite{Buckley:2010ar}, 
on {\tt HEPdata}~\cite{hepdata} 
and/or collaboration twiki pages~\cite{Atlastwiki,Cmstwiki}.  
The {\tt inSPIRE}~\cite{inspire} project may help to build a coherent information system, 
with detailed searchable and citeable entries.  
The ultimate goal should be to store all analysis information systematically 
in a common public archive based, e.g., at CERN. This brings us to 

\begin{quote}
{\bf Recommendation~\arabic{recom}b:} \emph{\databaseRec}
\end{quote}

\noindent

The analysis database should also be capable of storing any analysis-related software that may be provided alongside the analysis.  Although this is not listed as a recommendation, it would be extremely useful, and elegant, to design the infrastructure of analyses in a highly modular fashion so that cuts that define event
selections, or perhaps even object definitions, or codes that perform kinematic reconstructions, or that compute the variables on which an analysis is based, are all encapsulated in well-defined functions that are independent of the provenance of the data they use.  In fact, several such functions decoupled from the internal software
infrastructure are already being made public by various analyses, and these functions can be added to the analysis database systematically in the future.  In the case of complex analyses such as multi-variate
analyses (MVAs), details about the MVA functions can help validate the tools developed by the user before applying them to models other than those used in the published analysis.

As mentioned, {\tt Rivet} and {\tt HEPdata} provide examples
of such a platform, possibly supported by the {\tt inSPIRE} indexing and
searching infrastructure. Their functionality could be adapted to
accommodate further needs, emerging from the discussions on the
implementation of Recommendation~1b. The continued development of such
tools should be encouraged, and could benefit from the support of
initiatives such as the LHC Physics Centre at CERN (LPCC).

\subsection{Detector Modeling}
\addtocounter{recom}{1}

\paragraph{A) Efficiency maps:}
Analyses often use different definitions of analysis objects. For
example, the definition of a candidate electron in one analysis may
use a different definition of isolation than in another, or one
analysis may use a cut on an MVA function to define an electron
candidate while another applies cuts to several measured quantities.
A well-understood way to shield a potential analyst from unnecessary
complexity is to provide efficiency maps for each candidate object.
Indeed experiments do provide efficiency maps along with some
analyses, and we strongly encourage this practice.  
For a reliable use of efficiency maps,  
\begin{itemize}
\item the definition of the object for which an efficiency is
provided, e.g. an offline isolated electron, missing energy trigger,
etc., 
\item the definition of the efficiency, e.g. in which fiducial volume,
or after which cuts an efficiency is defined,
\item the final state topology for which an efficiency is defined 
\end{itemize}
should be given precisely. Furthermore, it is very helpful if  
the efficiencies are presented in formats that can be
implemented easily, such as lists of numbers or mathematical functions
or standard digitized forms that are easy to interface with
simulators or analysis codes. We thus arrive at 

\begin{quote}
{\bf Recommendation~\arabic{recom}a:} {\emph \efficiencyRec}  
\end{quote}
These standard electronic forms could rely on a platform similar to
that discussed under Recommendation~1b, for example {\tt Rivet/HEPdata} data and
routines.

\paragraph{B) Public fast detector simulation:}

A fast detector simulation provides an approximate mapping from the
pre-detector data to the post-reconstruction data. Publicly available
fast detector simulations exist, like {\tt PGS}~\cite{PGS} or 
{\tt  DELPHES}~\cite{Ovyn:2009tx}, which perform quite well and are
generally found to reproduce ATLAS and CMS results with reasonable
precision. Continued development, support and validation of these
tools is of high value.

\begin{quote}
{\bf Recommendation~\arabic{recom}b:} {\emph \simulatorRec}
\end{quote}

We propose an open Workshop, bringing together the developers of the
existing tools, the experts from the experiments, and the potential
users, to discuss possible means of implementing this recommendation.

For completeness we note that publishing unfolded results provides an
approach alternative to the need for processing MC data through a
detector simulation. Unfolding and correcting to the particle level is
certainly a preferred approach in many cases, e.g.\ in the presentation
of cross-sections or distributions, but in practice it is not always 
viable or desirable in the case of BSM searches.

\subsection{Analysis dissemination}
\addtocounter{recom}{1}

\paragraph{A) Basic requirements:}

It is extremely important that all the crucial numbers regarding
experimental results be summarized in a clear, concise, yet complete manner, preferably in tables.  
Experimental publications routinely provide these numbers, 
nevertheless here we encourage the maintenance of this good practice and list the minimum
required information.

\begin{quote}
{\bf Recommendation~\arabic{recom}a:}
   \emph{\analysisNumbersRec} 
\end{quote}

\noindent
In the case of a single-bin counting experiment these numbers include:
\begin{itemize}
\item $D$ --- the number of observed events in the signal region,
\item $\delta S$ --- the systematic error on the expected number of signal
  events across the parameter space of the new physics model considered,
\item $B$ --- the background estimate,
\item $\delta B$ --- the estimated background uncertainty, 
\item $\cal{L}$ --- the integrated luminosity estimate, and 
\item $\delta\cal{L}$ --- the integrated luminosity uncertainty.
\end{itemize} 
If the background estimate is the result of extrapolating from a control
region (e.g, from a side-band) to the signal region, the following should also
be provided (perhaps in the auxiliary information): 
\begin{itemize}
\item $Q$ --- the observed number of events in the control region
\item $\delta Q$ --- the uncertainty in Q
\item $k$ --- the ratio of expected background in the control region to the
  expected  background in the signal region.
  If the uncertainty $\delta k$
  on $k$ is not negligible, it should also be included. 
\end{itemize}
In the case of a multi-bin analysis, the above numbers should be given for each bin.
 
An important complication to address is how to account for
systematic uncertainties. For the single-bin analysis, 
numbers should be reported with and without the inclusion of systematic
uncertainties. 
The same holds for theoretical uncertainties of various types: it would be useful if the experiments
also provided results obtained 
without the inclusion of the theoretical uncertainties for well-specified theoretical 
inputs (such as parton distribution functions (PDFs), top mass, etc.). 
In particular, since theoretical uncertainties are not static, this has the advantage of facilitating 
their re-assessment at a later stage in a straightforward manner. 
Systematic uncertainties on the signal, $\delta S$, should be given
separately for detector specific sources, and for SM theory
uncertainties, such as PDFs. The systematic uncertainty stemming 
purely from the calculation of the signal model prediction should be
left out.

Whilst this method suffices for a single-bin analysis, it is not adequate for 
multiple-bin analyses, because it does not account for statistical dependencies between bins. 
One way to account for the lowest-order statistical dependencies, i.e., linear correlations, is to 
provide the correlation matrix. However, this approach breaks down if the uncertainties are large 
or if errors are highly asymmetric. 

Since it is common practice to include systematic uncertainties by integrating
over systematic parameters, we include the following recommendation, which
provides a straightforward way to perform this integration (or to profile). 

\begin{quote}
{\bf Addendum to  \arabic{recom}a:}
\emph{\multiBinRec}  
\end{quote}

\paragraph{B) The full likelihood:}

The statistical model of an analysis provides the complete mathematical
description of that analysis. 
The statistical model, through the probability density $p(o|x)$, relates the
observed quantities $o$ to the parameters $x$,  
describing the prediction in a model-independent way. 
By definition, the likelihood for a given set of theoretical model parameters $x$
is the   probability density over the observables $o$
evaluated at their observed values $O$. (For clarity,  
we denote observed quantities, e.g. $O$, by upper case symbols and parameters,
e.g. $o, x$, by lower case symbols.)  

The likelihood is the appropriate starting point for any detailed
interpretation of experimental results. However, many published 
analyses use likelihoods implicitly rather than explicitly. One problem
with using likelihoods implicitly (for example, when 
results are expressed as $O \pm \delta O)$ is that possible non-Gaussian tails are
ignored.  
If the uncertainties are small this is not an issue. However, if the
uncertainties are large
the likelihood should be modelled explicitly. Given the likelihood, all the standard
statistical approaches are available for extracting information from it. We therefore
suggest the following. 

\begin{quote}
{\bf Recommendation~\arabic{recom}b:}
\emph{\likelihoodRec} 
\end{quote}

\noindent
Often, likelihood functions are constructed in several steps involving several
hierarchies of fitted functions. Here, we define the \underline{final}
likelihood function as the last step in this process: it may be expressed in
terms of an integral or maximisation over the product of several (possibly
fitted) functions, which may be Gaussian or Poisson distributed, for example. 

Since providing the full likelihood 
requires a very good understanding of its  construction as well as of how to make correct 
use of it, a first approach towards Recommendation~\arabic{recom}b may be to provide 
likelihoods in an approximate (simplified) form, again with the range of validity---typically 
the region where the signal hypothesis is defined---clearly specified. 

It is, moreover, worth making the distinction between likelihoods parametrized only in terms of parameters that modify the normalization of distributions (e.g.\ cross-sections and branching ratios) versus those that parametrize acceptance (e.g.\ masses, spins, couplings of intermediate particles,  etc.).  The former is conceptually simple, as the distributions are trivially related to the parameters.  The latter is very difficult, since the distributions depend non-trivially on the parameters.  Parametrizing acceptance over model parameters often requires interpolation or even extrapolation, which are both a legitimate cause for concern.  Clearly, it is up to the experiments to make the judgement as to which parameters they feel the likelihood can be properly parametrized.

It would also be very useful and practical if the likelihood was provided in addition 
in a digital form.  There already exists a generic, unified
framework, {\tt RooStats}~\cite{Moneta:2010pm}, used 
by many LHC analyses, which allows one to model the probability density
functions and likelihoods required as an input for any statistical
inference technique, and also provides a set of major statistical
techniques as {\tt c++} classes with coherent interfaces to the statistical model.  
Publication of likelihoods in a systematic fashion under a standard digital format
would also make combination of results much more feasible.

\begin{quote}
{\bf Recommendation~\arabic{recom}c:}
\emph{\roostatsRec}  
\end{quote}

We also note at this point that the {\tt RECAST}~\cite{Cranmer:2010hk} project would allow one to obtain the signal contribution to the likelihood for an arbitrary theoretical model, thus allowing one to build a higher-level framework for analysis re-interpretation.

\subsection{Interpretation of experimental results}
\addtocounter{recom}{1}

So far our recommendations concern generally the presentation of experimental results, irrespective 
of whether they report a signal or are used to set exclusion limits. 
Let us now turn to the interpretation of these results, the presentation of confidence 
intervals, parameter inference and limit setting.
 
Many different forms of experimental limit exist. Commonly, one-sided
limits are derived in the absence of a signal observation, as is currently the case, 
but this will switch to two-sided limits (constraints) in case of a discovery.  
Limits may be quoted in various different schemes (such as 
Feldman-Cousins, $\mathrm{CL}_s$, etc). 
It is crucial that the limit setting procedure be explicitly  
defined in order to permit an informed comparison
of the quoted confidence level. 

The {\em shape} (steepness) of the confidence level is essential information, e.g., for analyses that combine different experimental searches. 
It is therefore important that constraints are shown at several, rather than just one, confidence levels.
Moreover, for the correct statistical interpretation,  the 
expected constraints should be given in addition to the observed ones.
Of course a more informative option would be to here, too, implement Recommendation~3b  
and publish the final likelihoods. 

Regarding uncertainties, as mentioned earlier, it would be useful if confidence intervals
were (also) presented for fixed input PDF's and other theoretical input, all
explicitly tabulated. Moreover, when the interpretation of the experimental results 
is done  in a ``model-independent'' way in terms of $\sigma\times {\rm BR}\times{\rm acceptance}$, 
the modeling of the acceptance should be precisely described.
We sum this up as 

\begin{quote}
{\bf Recommendation~\arabic{recom}:}
\emph{\interpretRec}
\end{quote}

\noindent
Note that Recommendation~\arabic{recom} in principle applies to any (re-)interpretation study, irrespective 
of whether it is done by an LHC collaboration or by non-collaboration groups. 
Needless to say, the model under investigation must be precisely defined. 

As an aside we note that when conducting searches for supersymmetry
or other new physics, experimental collaborations often use grids of models for
which signal cross-sections, acceptances, efficiencies, and exclusions are
evaluated, and then used to set limits by interpolation. It would be 
useful if these grid models were documented and fully specified in terms of
model inputs, spectrum information (e.g., SLHA files), predicted signal
cross-sections, acceptance$\times$efficiency after selections and cuts,
etc.. Also, since the tools provided by theorists constantly evolve, it is useful
to  document which tools and versions thereof have been used.

\subsection{Higgs Searches}
\addtocounter{recom}{1}

Given the special role of Higgs searches, we make a specific and separate recommendation for them. 
Higgs bosons are searched for in many different
possible topologies, each of which are predicted to be present at some level,
dictated by Higgs branching ratios in the Standard Model or new physics
model. 
Many Higgs searches may be interpreted
within the Standard Model itself, but both the branching ratios and the
production cross-sections  and distributions 
(and indeed the number of Higgs particles)
 may differ in new physics models. For this
reason, it is important to include the channel-by-channel information in
the primary Higgs search papers. Of course this does not preclude an additional
combination of channels assuming some model, e.g., the Standard Model.

\begin{quote}
{\bf Recommendation~\arabic{recom}:}
\emph{\higgsRec}
\end{quote}

Many publications on Higgs searches are already consistent with this recommendation in that constraints
for individual production/decay modes are presented.  The procedure of doing so as a function of the Higgs mass is crucial, especially in the context of multiple (or composite) Higgs boson models. 
Indeed, different Higgs models weight various possible production mechanism and decay distributions differently.  
It is moreover very instructive to give the best-fit signal strengths, $\sigma/\sigma_{\rm SM}$, as function of the SM Higgs boson mass for all available channels, along with error bands, 
as this facilitates testing deviations from SM couplings (see e.g.\ the discussion in  \cite{Espinosa:2012ir}).

In order to make the channel-by-channel data available
for an analysis of the Higgs sector of (nearly any) arbitrary model, 
it is also important that  the expected sensitivity be reported for each channel 
in addition to the observed data. 
This permits the selection of the potentially strongest search channel, 
and allows to retain the correct statistical interpretation of, e.g.,  
95\% C.L.\ exclusion bounds~\cite{Bechtle:2011sb}. 
Finally, in setting
limits or analyzing positive signals it is crucial to give details of
acceptance systematics related thereto. These aspects are covered by
Recommendation~4. 
Needless to say, all of the other Recommendations also apply.

\subsection{Exclusive Analysis Design \label{sec:disjoint}} 
\addtocounter{recom}{1}

It is a common approach to confront a new physics model with multiple analyses.
In such cases, a likelihood is assigned to each analysis separately, and
then all likelihoods are combined. 
This combination is typically easy only if the statistical data from each
analysis are independent: then, one may combine the analyses
by simply taking the product of all likelihoods.  The correlations that may
arise in the systematic errors can be dealt with by following the addendum to
Recommendation~3a.  We realize that it is not possible to build an
experimental search program that consists fully of disjoint analyses, and
understand that avoiding overlaps between various inclusive analyses, e.g.,
those based on complicated kinematic variables, is far from trivial.  Our
recommendation applies to simpler cases, e.g. when building analyses 
based on simple variables like object multiplicity, or when defining different search
regions in a single analysis.  It is our intend to emphasize
that more information is typically available in the combination of multiple
disjoint analyses than via a single analysis.   

\begin{quote}
{\bf Recommendation~\arabic{recom}: }
\emph{\designRec} 
\end{quote}

\clearpage
\section{Executive Summary of Recommendations}

We here summarize our recommendations. We remind the reader that whenever we 
split into several steps, options   
{\bf (a)} should be understood as ``crucial" recommendations, while 
{\bf (b), (c)} are ``desirable steps''. 
For completeness we also note that the ordering of Recommendations 1--6 does not imply prioritizing. \\

\begin{enumerate}
\item \begin{enumerate} 
      \item \describeRec \vskip 0.2cm
      \item \databaseRec 
      \end{enumerate}
\vskip 0.4cm
\item \begin{enumerate} 
      \item \efficiencyRec \vskip 0.2cm 
      \item \simulatorRec 
      \end{enumerate}
\vskip 0.4cm
\item \begin{enumerate} 
      \item \analysisNumbersRec \\
      \emph{Addendum}:\\ \multiBinRec \vskip 0.2cm
      \item \likelihoodRec \vskip 0.2cm
      \item \roostatsRec
      \end{enumerate}
\clearpage
\item\interpretRec
\vskip 0.4cm
\item \higgsRec
\vskip 0.4cm
\item \designRec
\end{enumerate}


\section{Conclusions}
This document presents a set of recommendations for the presentation of LHC  
results on searches for new physics, which are aimed at providing a more efficient flow 
of scientific information and at facilitating the interpretation of the results in wide classes of models. 
It originated from discussions at the Les Houches ``Physics at TeV Colliders 2011'' 
workshop~\cite{LHproc} and was thoroughly discussed and refined, with valuable input from 
representatives of the ATLAS and CMS collaborations, in a dedicated miniworkshop organized 
by the LHC Physics Centre at CERN~\cite{feb13}. 
The target of
these recommendations are physicists both within and outside the LHC
experiments, interested in the best exploitation of the BSM search analyses.  

The added value for the experiments, and the whole HEP community, in
extending the scope of the information made available about the
experimental results, is a faster and more precise feedback
on the implications of these results for a broad range of theoretical
scenarios. Correlations and consistency checks among the findings of
different experiments, at the LHC and elsewhere, will be facilitated,
and will provide crucial input in the choice of the best research
directions in both the near and far future, at the LHC and
elsewhere. Improving the way the results of the LHC searches are
documented and stored furthermore provides a forum to
explore alternative approaches to long-term data archiving. 

The tools needed to provide extended experimental information will
require some dedicated efforts in terms of resources and manpower, to
be supported by both the experimental and the theory
communities. Practical solutions towards the development of these
tools and the implementation of the proposed recommendations will be
addressed in dedicated workshops and working groups.

\section*{Acknowledgements}
We are grateful to the ATLAS and CMS search group conveners for helpful and supportive advice. 
We are also grateful to the ATLAS and CMS physics coordinators, in particular Richard Hawkings and Greg Landsberg, for very constructive discussions on these recommendations.  
BCA~would like to thank other members of the Cambridge SUSY Working Group for
helpful suggestions. 

This work has been partially supported by IN2P3, 
the Royal Society, STFC, and the ARC Centre of Excellence for Particle Physics at the Terascale.
The work of JE is supported in part by the London Centre for Terauniverse Studies (LCTS), using funding from the European Research Council 
via the Advanced Investigator Grant 267352. 
JFG is supported by US DOE grant DE-FG03-91ER40674.  
HBP is supported in part by US DOE grant DE-FG02-97ER41022..



\end{document}